\documentclass[twocolumn,secnumarabic,amssymb, nobibnotes, aps, prd]{revtex4}

\usepackage{color}
\usepackage{pstricks}
\usepackage{graphicx}
\usepackage{amsmath}

\newcommand{\bit}{\begin{itemize}}
\newcommand{\eit}{\end{itemize}}
\newcommand{\ben}{\begin{equation}}
\newcommand{\een}{\end{equation}}
\newcommand{\bea}{\begin{eqnarray}}
\newcommand{\eea}{\end{eqnarray}}
\newcommand{\bens}{\begin{equation*}}
\newcommand{\eens}{\end{equation*}}
\newcommand{\beas}{\begin{eqnarray*}}
\newcommand{\eeas}{\end{eqnarray*}}

\newcommand{\p}{\partial}

\begin{document}

\title{$D7$-$\overline{D7}$ bilayer:  holographic 
dynamical symmetry breaking}

\author{Gianluca Grignani$^1$, Namshik Kim$^2$, Gordon W. Semenoff$^2$}
\affiliation{1) Dipartimento di Fisica, Universit\`a di Perugia,
I.N.F.N. Sezione di Perugia,
Via Pascoli, I-06123 Perugia, Italy}
\affiliation{2) Department of Physics and Astronomy, University of British
Columbia, Vancouver, British Columbia, Canada V6T 1Z1}                                                                                                                                                                                                                                     

\begin{abstract}
We consider a holographic model of dynamical symmetry breaking in 2+1-dimenisons,
where a parallel D7-anti-D7
brane pair fuse into a single object, corresponding to the $U(1)\times U(1)\to U(1)$ symmetry breaking
pattern.  We show that the current-current correlation functions
can be computed analytically and exhibit the low momentum structure that is expected when 
global symmetries are spontaneously broken.  We also find that these correlation functions
have poles attributable to infinite towers of vector mesons with equally spaced masses.
\end{abstract}

\maketitle

In weakly coupled quantum field theory, spontaneous symmetry breaking
is a familiar paradigm.  It is based on formation of a condensate,
usually an order parameter obtaining a nonzero expectation value and
the resulting features of the spectrum such as goldstone bosons and a
Higgs field.  String theory holography has given an alternative
picture of dynamical symmetry breaking in terms of geometry.
Particularly with probe branes, the symmetry breaking corresponds to
the branes favoring a less symmetric worldvolume geometry over a more
symmetric one.  This is seen in the Sakai-Sugimoto model of
holographic quantum chromodynamics \cite{Sakai:2004cn}.  There, chiral
symmetry breaking corresponds to the fact that a $D8$-$\overline{D8}$ brane
pair prefer to fuse into a cigar-like geometry, rather than remaining
in a more symmetric independent configuration.  
In this paper, we shall study a model which is close in spirit to the
Sakai-Sugimoto model, the $D7$-$\overline{D7}$ system which has a
2+1-dimensional overlap with a stack of $D3$-branes.  It can be
considered a toy model of chiral symmetry breaking in strongly coupled
2+1-dimensional quantum field 
theories containing fermions and it is explicitly solvable.
The symmetry breaking pattern is $U(N)\times U(N)\to U(N)$ and, at
least in principle, it is possible to gauge various subgroups of the
global symmetry group and to study the Higgs mechanism at strong coupling. In
the following we shall concentrate on the case $U(1)\times U(1)\to
U(1)$ which displays the essential features of the mechanism.

{\begin{figure}
\includegraphics[scale=.15]{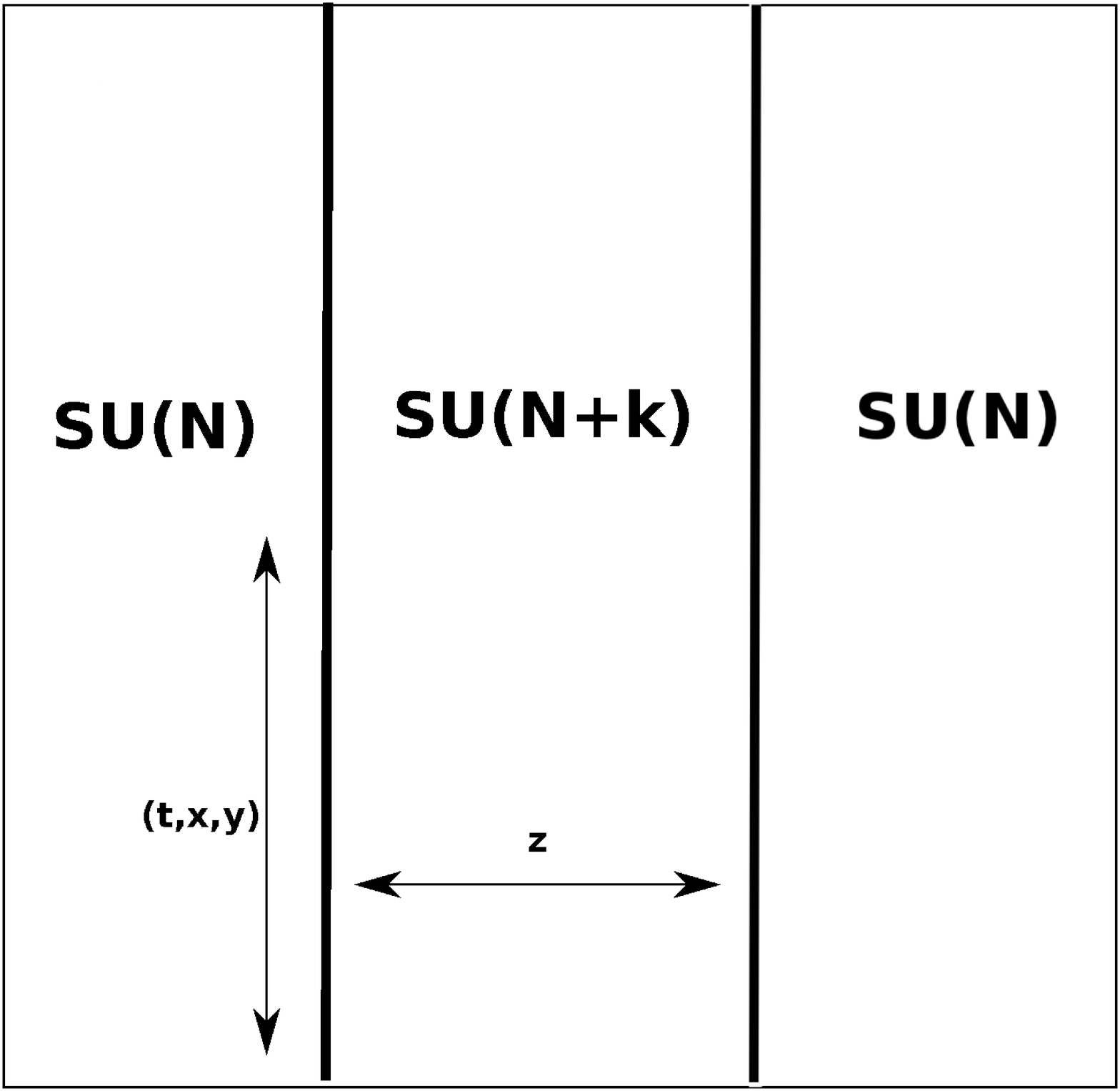}~~~~~~~~
\begin{caption} {Two parallel 2-dimensional spaces, depicted by the 
vertical dark lines, are inhabited by fundamental representation
fermions which 
interact via fields of ${\mathcal N}=4$ supersymmetric Yang-Mills theory
in the bulk.  The Yang-Mills theory in the region between
the layers has a different rank gauge group than that in the regions
external to the bilayer.  
\label{bilayer} }\end{caption}
\end{figure}

Before analyzing the $D7$-$\overline{D7}$ system, let us discuss its
quantum field theory dual, the bilayer system depicted in figure
\ref{bilayer}.  Massless relativistic 2+1-dimensional fermions are
confined to each of two parallel but spatially separated layers.  They
are two-component spinor representations of the SO(2,1) Lorentz group
with a U(1) global symmetry for the fermions inhabiting each layer.
The overall global symmetry is thus $U(1)\times U(1)$.  The
3+1-dimensional bulk contains ${\mathcal N}=4$ supersymmetric
Yang-Mills theory.  The fermions transform in the fundamental
representation of the gauge groups of the Yang-Mills theories.  As
shown in figure 1, the rank of the Yang-Mills gauge groups differ in
the interior and exterior of the bilayer by an integer $k$ which
arises from the worldvolume flux in the $D7$-$\overline{D7}$
system.  The D-brane system which we
shall discuss studies this theory in the strong coupling planar limit
where, first, the Yang-Mills coupling $g_{\rm YM}$ is taken to zero
and $N$ to infinity while holding $\lambda\equiv g_{YM}^2 N$ fixed
and, subsequently, a strong coupling limit of large $\lambda$ is
taken. The field theory mechanism for the symmetry breaking which we
shall analyze is an exciton condensate which binds a fermion on one
layer to an anti-fermion on the other layer and breaks the $U(1)\times
U(1)$ symmetry to a diagonal $U(1)$.

{\begin{figure}
\includegraphics[scale=.14]{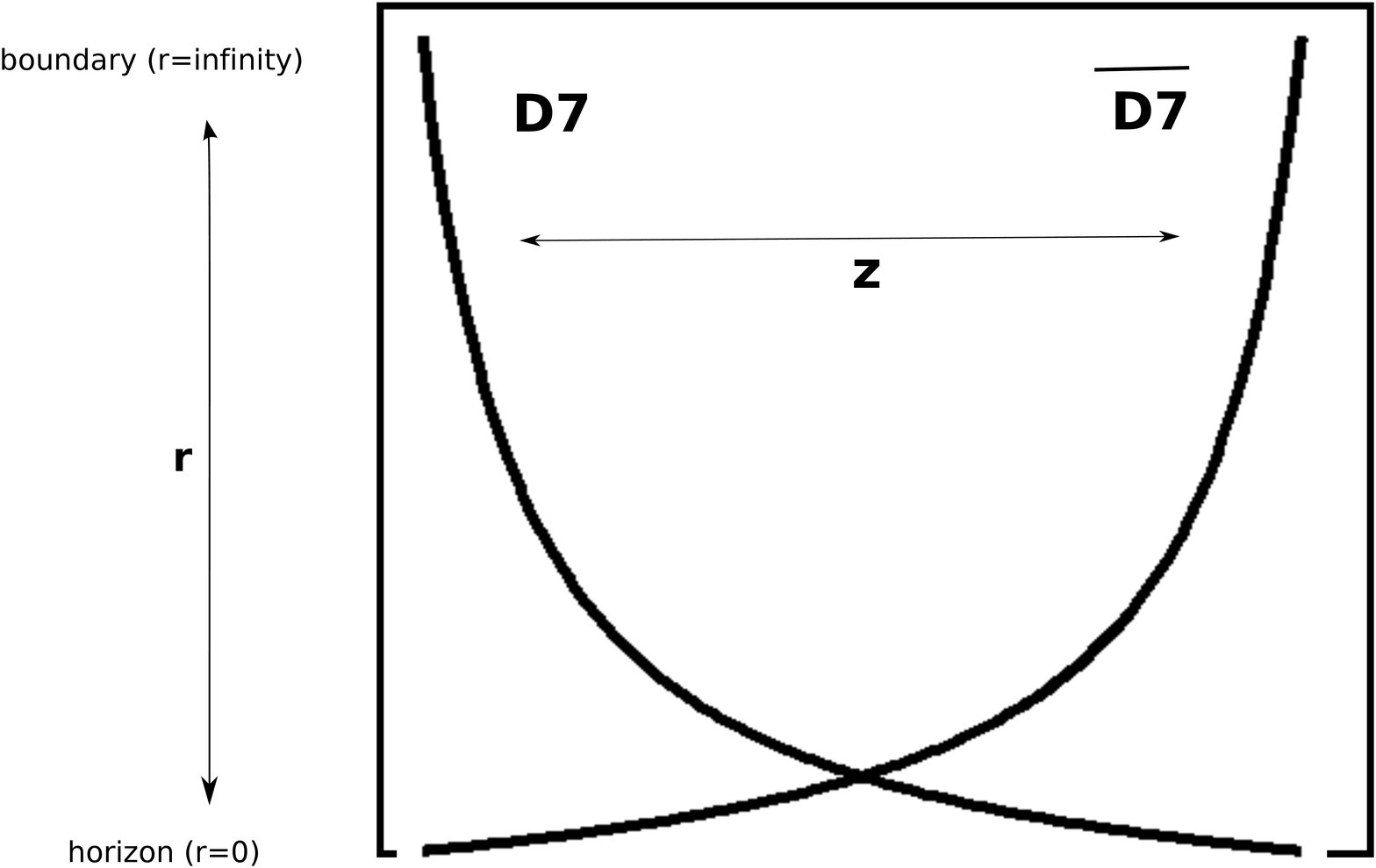}~~~
\begin{caption} { The z-position of the D7-branes depends on AdS-radius and
with the appropriate orientation the branes would always intersect.
  \label{d7d7bar}
}\end{caption}
\label{fig:intersecting}
\end{figure}

There has been significant recent interest in graphene bilayer systems
where formation of an exciton driven dynamical symmetry breaking of
the kind that we are discussing has been conjectured \cite{macdonaldII}. The geometry is
similar, with the layers in figure \ref{bilayer} replaced by graphene
sheets and the space in between with a dielectric insulator.  In spite
of some differences: graphene is a relativistic electron gas with a
strong non-relativistic Coulomb interaction, whereas what we describe
is an entirely relativistic non-Abelian gauge theory, there are also
similarities and perhaps lessons to be learned. For example, we find
that the exciton condensate forms in the strong coupling limit even in
the absence of fermion density whereas the weak coupling computations
that analyze graphene need nonzero electron and hole densities in the
sheets to create an instability. We also find ``coulomb drag'', where
the existence of an electric current in one layer induces a current in
the other\cite{Falko}.  In the holographic model, the drag would vanish in the
absence of a condensate, whereas it is large when a condensate is
present. The correlator between the electric currents≈ß in the two sheets
(from (\ref{bilayeraction}) below) is
\begin{align}\label{grapheneinterlayer}
<j_a(k)\tilde j_b(\ell)>=   
\frac{4\lambda(1+f^2)|k|}{(2\pi)^2\sinh2|  k| \rho_m}\left(\delta_{ab}-\frac{ k_a k_b}{ k^2}\right)\delta(k+\ell)
\end{align} where $|k|=\sqrt{\vec k^2-\omega^2/v_F^2}$, there is a factor of $4$ from
the degeneracy of graphene, $v_F$ is the electron fermi velocity and
$\lambda$ and $ f^2$ are parameters and $\rho_m$, given in
(\ref{rhom}), is proportional to the interlayer spacing.  Aside from the superfluid pole at $k^2=0$, this
correlator has an infinite series of poles at $k^2=(n\pi/\rho_m)^2$,
$n=1,2,...$ due to vector mesons. Parameters partially cancel in the
ratio of the current-current correlator in (\ref{grapheneinterlayer})
to the single layer correlator, $<j\tilde j>/<jj>={\rm csch}2|k|\rho_m$.

{\begin{figure}
\includegraphics[scale=.14]{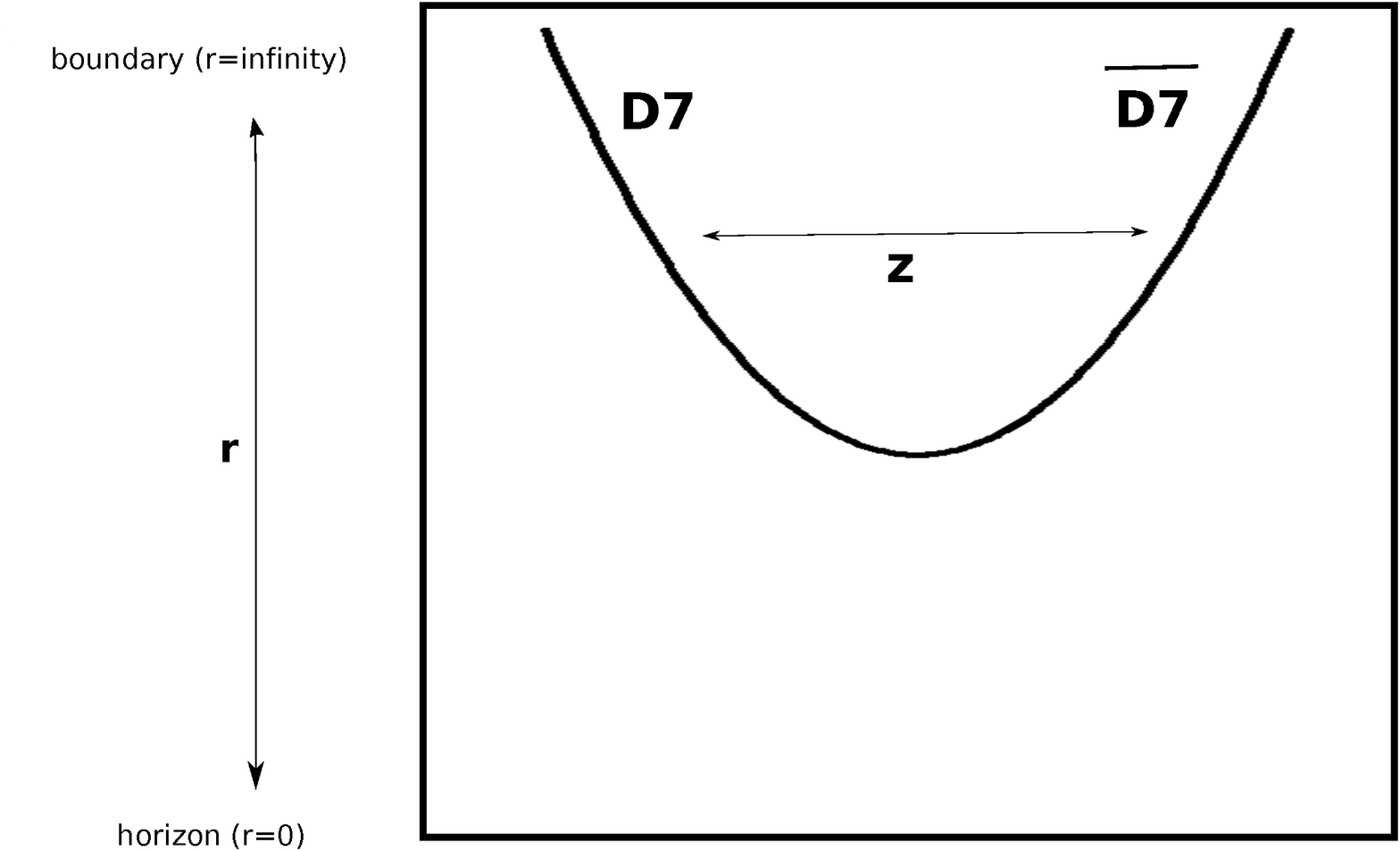}\\
\begin{caption} {Joined configuration.  \label{d7d7barjoin}
}\end{caption}
\end{figure}

Symmetry breaking in the $D7$-$\overline{D7}$ system has already been
studied in reference \cite{Davis:2011am}.  The mechanism is a joining
of the $D7$ and $\overline{D7}$ worldvolumes as depicted in figure
\ref{d7d7barjoin}.  The $D7$ and $\overline{D7}$ are probe branes \cite{Karch:2000gx} 
in the $AdS_5\times S^5$ geometry which is the holographic dual of
3+1-dimensional ${\mathcal N}=4$ supersymmetric Yang-Mills theory.
A single probe D7-brane is stable when it has magnetic flux added
to its worldvolume \cite{Bergman:2010gm}. Its most symmetric
configuration is dual to a defect conformal field theory
\cite{Bergman:2010gm}\cite{Davis:2011gi} where the flux ($f$ in the
following) is an important parameter which determines, for example,
the conformal dimension of the fermion mass operator.  The
$D7$-$\overline{D7}$ pair would tend to annihilate and are prevented from
doing so by boundary conditions that contain a pressure (the parameter
$P$ in the following) which holds them apart.  The problem to be solved 
is that of finding the configuration of the $D7$ and $\overline{D7}$ in
the $AdS_5\times S^5$ background, subject to the appropriate boundary
conditions.  We shall impose the parity and time-reversal invariant
boundary conditions that were discussed in reference
\cite{Davis:2011gi}.  We differ from reference \cite{Davis:2011am} in that
we use the zero temperature limit, a simplification that  
allows us to obtain our main result, explicit current-current correlation functions
for the theory described by the joined solution (\ref{bilayercurrents1})-(\ref{bilayercurrents}). 
The $AdS_5\times S^5$ metric is
\begin{align}
ds^2=R^2[r^2(-dt^2+dx^2+dy^2+dz^2)+\frac{dr^2}{r^2}\nonumber \\
+d\psi^2 + \sin^2\psi d\Omega_2^2 + \cos^2\psi d\tilde\Omega_2^2]
\label{adsmetric}
\end{align}
where 
$d\Omega_2^2$ and $
d\tilde\Omega_2^2$ are metrics of unit 2-spheres and
$\psi\in[0,\frac{\pi}{2}]$.  The radius of curvature is
$R^2=\sqrt{4\pi g_sN}{\alpha'}$, where $g_s$ is the closed string
coupling constant and $N$ the number of units of Ramond-Ramond 4-form
flux of the IIB string background.  The holographic dictionary sets
$g_{\rm YM}^2=4\pi g_s$, and $N$ becomes the rank of the Yang-Mills
gauge group.  The embedding of the $D7$ in this space is mostly
determined by symmetry. We take the $D7$ and $\overline{D7}$ embeddings to
wrap $(t,x,y)$, $S^2$ and $\tilde S^2$ and to sit at the parity
symmetric point $\psi=\frac{\pi}{4}$.  To solve embedding equations,
the transverse coordinate $z$ must depend on the radius $r$. At the
boundary of $AdS_5$ ($r\to\infty$), we impose the boundary condition
that the $\overline{D7}$ is located at $z=-L/2$ and ${D7}$ at $z=L/2$.  The
worldvolume metric of one of the branes is then
\begin{align}
d\sigma^2 =R^2[r^2(-dt^2+dx^2+dy^2)+\frac{dr^2}{r^2}(1+r^4\dot z(r)^2)
\nonumber \\ +\frac{1}{2} d\Omega_2^2 + \frac{1}{2}\tilde d\Omega_2^2]
\label{D7metric}
\end{align}
where $\dot z=dz/dr$. The field strength of the world-volume gauge
fields are
\begin{align}
F= \frac{R^2}{2\pi\alpha'}\frac{f}{2}\Omega_2 + \frac{R^2}{2\pi\alpha'}\frac{f}{2}\tilde \Omega_2
\end{align}
where $\Omega_2$ and $\tilde \Omega_2$ are the volume forms of the
unit 2-spheres.  The flux forms two Dirac monopole bundles, each with
monopole number $n_D= \sqrt{\lambda}f^2$.  Stability and other
properties of the theory \cite{Bergman:2010gm}\cite{Davis:2011gi}
require that $23/50\leq f^2\leq 1$, otherwise it is a tunable parameter.
The embedding is determined by extremizing the Dirac-Born-Infeld plus
Wess-Zumino actions,
\begin{align}
S=-\frac{T_7N}{g_s}\int d^8\sigma\left[
\sqrt{-\det(g(\sigma)+2\pi\alpha' F)}  \right. \nonumber \\ \left. 
\mp\frac{(2\pi\alpha')^2}{2}F\wedge F\wedge C_{4}\right]
\label{dbi}
\end{align}
$T_7=1/(2\pi)^7{\alpha'}^4$ is the brane tension, $C_{4}$ is the
Ramond-Ramond 4-form of the IIB string background and the $\mp$ refer
to the $D7$ and $\overline{D7}$, respectively.  With our Ansatz, this
reduces to a variational problem with Lagrangian
\begin{align}
{\mathcal L}=(1+f^2)r^2 \sqrt{1+r^4\dot z(r)^2}\mp f^2
r^4\dot z(r)
\label{dbi2}
\end{align}
$z(r)$ is a cyclic variable
whose equation of motion 
is solved by
$z_\pm(r)=\pm\frac{L}{2}\mp\int_r^\infty dr \dot z_+(r)$
 is the position of the brane to the right (upper sign) or left (lower sign) of $z=0$
and 
$
\dot z_\pm(r)= \pm \frac{f^2r^4+ P}{r^2\sqrt{(r^4-P)((1+2f^2)r^4+P)}}
$.
$P$ is an integration constant proportional to the
pressure needed to hold the branes with their asymptotic separation
$L$.  When they are not joined, they do not interact, at least in this
classical limit, and $P$ must be zero.  Then
$z_\pm(r)=\pm\frac{L}{2}\mp\frac{f^2}{\sqrt{1+2f^2}~r}$ as depicted in figure
\ref{d7d7bar}.  When they are joined, as depicted in figure
\ref{d7d7barjoin}, $P$ must be nonzero and they are joined at a minimum radius
$r_0=P^{\tfrac{1}{4}}$ and $L$ and $P$ are related by $
LP^{\frac{1}{4}}=2\int_{1}^\infty dr\frac{f^2r^4+
  1}{r^2\sqrt{(r^4-1)((1+2f^2)r^4+1)}} $.

The joined solution will always be the lower energy solution when the
branes are oriented as in figures \ref{d7d7bar} and
\ref{d7d7barjoin}. They are also stable for any value of $L$ when the
brane and antibrane are interchanged, the ``chubby solutions''
discussed in reference \cite{Davis:2011am}, only when $23/50\leq
f^2\lesssim .56$.  When $f^2>.56$ the chubby solutions are unstable for
any $L$.  (As noted in reference~\cite{Davis:2011am}, there can be a
much richer phase structure when temperature, density or external
magnetic fields are introduced.) For the chubby solution, the gauge group ranks
$N$ and $N+k$ in figure \ref{bilayer} trade positions. 

A simple diagnostic of the properties of the fermion system in the
strongly coupled quantum field theory which is dual to the joined
branes is the current-current correlation function.  It is obtained by
solving the classical dynamics of the gauge field on the world-volume
of the branes with the Dirichlet boundary condition.  The quadratic
form in boundary data in the on-shell action yields the
current-current correlator.  Here, the brane geometry is simple enough
that, to quadratic order, AdS components of the vector field decouple
from the fluctuations of the worldvolume geometry, as well as from
those components on $S^2,\tilde S^2$. To find them, we simply need to
solve Maxwell's equations on the worldvolume,
$$
\partial_B \left[\sqrt{g}g^{BC}g^{DE}(\partial_C A_E
-\partial_E A_C)\right]=0
$$ where the worldvolume metric is given in equation \eqref{D7metric}
above and the gauge fields have indices $B,C,...=(t,x,y,r)$.  In
the $A_r=0$ gauge,
\begin{align}
\partial_r (\partial_a 
A_a)=0
~~,~~
\partial_\rho^2 A_a +
\partial_b(\partial_b A_a
-\partial_a A_b)=0
\end{align}
where indices $a,b,...=(t,x,y)$, we have suppressed the Minkowski metric 
for contracted indices and we have redefined the radial coordinate as
$
\rho = \int_r^\infty \frac{dr}{r^2}\sqrt{1+r^4\dot z^2}
$.
In the simpler case of a single D7-brane, say the brane which originates 
on the right in figure \ref{d7d7bar}, 
whose geometry is $AdS_4$, these equations are solved by \cite{Davis:2011gi} 
$$
A_a(k,\rho)= A_a(k)\cosh|k|\rho +\frac{1}{|k|} A_a'(k)\sinh|k|\rho 
$$ where $A_a(k,\rho)=\int d^3x e^{ikx}A_a(x,\rho)$, $k_a
A_a(k)=0=k_a A_a'(k)$ and $|k|=\sqrt{\vec k^2-k_0^2}$. Regularity
at the Poincare horizon ($\rho\to\infty$) requires $A_a'(k)=-|k| A_a(k)$.  Moreover, with
the on-shell action,
$$
S=
-\frac{N(f^2+1)}{4\pi^2}\int d^3k |k| A_a(-k)\left(\delta_{ab}-k_ak_b/k^2\right)A_b(k)+\ldots
$$
$e^{-S}$ is a generating function for current-current correlators in the dual conformal field 
theory where the U(1) symmetry is global, ($j_a(k)=g_{\rm YM}\delta/\delta A_a(-k)$)
\begin{align}\label{intralayercurrentcurrent}
 <j_a(k)j_b(\ell)>= \frac{\lambda(f^2+1)}{2\pi^2}
|k|\left(\delta_{ab}-k_ak_b/k^2\right)\delta(k+\ell)
\end{align} 
Alternatively, if instead of the Dirichlet boundary conditions used
above, we impose the Neuman boundary condition that $\partial_\rho
A_a(k,\rho)$ approaches $ A_a'(k)$ as $\rho\to0$, we can write
the on-shell action as a functional of $A'(k)$ and 
it generates correlators of the gauge field in a different
conformal field theory where the U(1) symmetry is gauged and the gauge
field is dynamical. It yields the Landau gauge 2-point function of the
photon field in that theory \cite{Marolf:2006nd} ($a_a(k)=\delta/\delta A'(-k)$),
$$
<a_a(k) a_b(\ell)>=\frac{N(f^2+1)}{2\pi^2}
\frac{1}{|k|}\left(\delta_{ab}-k_ak_b/k^2\right)\delta(k+\ell)
$$ The momentum dependence of these correlation functions is
consistent with conformal symmetry.  

To analyze the joined configuration, we note that in that case $\rho$ reaches
a maximum
\begin{align}\label{rhom}
\rho_m = \frac{L}{2}\frac{\int_0^1 \frac{dx (1+f^2)}{\sqrt{(1-x^4)((1+2f^2)-x^4)}}}
{\int_0^1\frac{dx(f^2+x^4)}{\sqrt{(1-x^4)((1+2f^2)-x^4)}}}
\end{align}
We use a variable $s=\rho$ for the left branch and $s=2\rho_m-\rho$
for the right branch of figure \ref{d7d7barjoin}.  With the Dirichlet
boundary conditions $A_a(k,s=0)=A_a(k)$ and $A_a(k,s=2\rho_m
)=\tilde{A}_a(k)$ the on-shell action is
\begin{align}\tilde{S}&=-\frac{N(f^2+1)}{4\pi^2} \int d^3k
\left[  (|A_a(k)|^2+|\tilde A_a(k)|^2)\coth 2|k|\rho_m\right.\nonumber\\
&\left.
-2
A_a(-k)\tilde A_a(k){\rm csch}2|k|\rho_m\right]+\ldots
\label{bilayeraction}\end{align}
The current-current correlation functions can are
diagonalized by 
$
j_+\equiv j+\tilde{j}
$, 
$j_-\equiv j-\tilde{j}
$, so that
\begin{align}
<j_{a+}j_{b-}>&=0 \label{bilayercurrents1}\\
<j_{a+}j_{b+}>
&=\frac{\lambda(f^2+1)}{2\pi^2}  k  \tanh  k \rho_m\left(\delta_{ab}-\frac{ k_a k_b}{ k^2}\right)
\label{bilayercurrents2}\\
<j_{a-}j_{b-}>
&=\frac{\lambda(f^2+1)}{2\pi^2}  k  \coth  k \rho_m\left(\delta_{ab}-\frac{ k_a k_b}{ k^2}\right)
\label{bilayercurrents}
\end{align}
At large Euclidean momenta, (\ref{bilayercurrents2}) and (\ref{bilayercurrents}) revert to the
conformal field theory correlators in (\ref{intralayercurrentcurrent}). At time-like momenta
the correlator $<j_{a-}j_{b-}>$ has a pole at $k^2=0$ which is the
signature of dynamical breaking of a diagonal $U(1)$ subgroup of the
$U(1)\times U(1)$ symmetry and gives rise to superfluid linear
response.  On the other hand, the correlator $<j_{a+}j_{b+}>\sim k^2$
for small $k$, which indicates that the system is an insulator in the
channel which couples to the other diagonal U(1) subgroup with current
$j_{a+}$.  In addition, both correlators have an interesting analytic
structure. They have no cut singularities.  $<j_{a+}j_{b+}>$ has poles
at the energies
\begin{equation}
\label{polep}
 k_0^2= k_1^2+ k_2^2+\left(\frac{\pi(2n+1)}{2 \rho_m}\right)^2\ ,~~~~ n=0,1,\dots
\end{equation}
and $<j_{a-}j_{b-}>$ has poles at 
\begin{equation}
\label{polem}
 k_0^2= k_1^2+ k_2^2+\left(\frac{\pi n}{ \rho_m}\right)^2\ ,~~~~ n=0,1,\dots
\end{equation}
indicating two infinite towers of massive spin-one particles.  These
would be narrow bound state resonances with decay widths that vanish
as $N\to \infty$, as one expects in the large-$N$ limit that we are
studying here ~\cite{Witten:1979kh}. The current operators create
these single-particle states from the vacuum.  Their creation of
multi-particle states, which would normally result in cut singularities, is
suppressed in the large $N$ planar limit.  The resonances are simply
the tower of vector mesons whose masses (\ref{polep}) and
(\ref{polem}) occur at eigenvalues of $-\partial_s^2$ with Dirichlet
boundary conditions on the interval $s\in[0,2\rho_m]$.  The fact that
currents create either even or odd harmonics is due to $L\to -L$
reflection symmetry.

In the above, we used Dirichlet boundary conditions for the
worldvolume gauge field. It is possible, alternatively, to select
Neumann boundary conditions by choosing $\p_sA_a$ rather than $A_a$ on
the asymptotic boundary.  The result is dual to a field theory where
the U(1) symmetries are gauged and the on-shell action generates
photon correlation functions~\cite{Marolf:2006nd}.  Most relevant
are mixed Neuman and Dirichlet boundary conditions.  
For example, in graphene, a diagonal electromagnetic U(1) is gauged whereas the
orthogonal U(1) is a global symmetry.  This is obtained by applying
the Dirichlet condition to $A(s=0, k)-A(s=2\rho_m, k)$ and the Neuman
condition to $\partial_sA(s=0, k)-\partial_sA(s=2\rho_m, k)$.  In this
case, the correlation functions are
\begin{align}
<j_aa_b>&=0\\
<j_aj_b>&=
\frac{\lambda(f^2+1)}{4\pi^2}   k  \coth  k  \rho_m\left(\delta_{ab}-\frac{ k_a k_b}{ k^2}\right)\\
<a_aa_b>&=\frac{N(f^2+1)}{4\pi^2} \frac{1}{ k } \coth  k  \rho_m\left(\delta_{ab}-\frac{ k_a k_b}{ k^2}\right)
\end{align}
The global U(1) symmetry is spontaneously broken and its current $j_a$
has a pole in its correlation function.  The unbroken gauged U(1) has a
massless pole corresponding to the photon.  In addition, the two towers of intermediate states
have the same masses with values (\ref{polem}). There is a family of
more general mixed boundary conditions which are interesting and which
will be examined in detail elsewhere.

This work is supported in part by NSERC of Canada and in part by the MIUR-PRIN contract 2009-KHZKRX.

\end{document}